\documentclass[a4paper,11pt]{article}
\usepackage{jheppub} % for details on the use of the package, please see the JINST-author-manual
\usepackage{lineno}
\usepackage{nicefrac}
\usepackage{etoolbox}
\usepackage{csquotes}
\makeatletter
\patchcmd{\thebibliography}{\advance\leftmargin\labelsep}{\leftmargin\z@\itemindent\z@}{}{}
\patchcmd{\@biblabel}{\hss}{\hfill}{}{}
\makeatother

\arxivnumber{2601.14394} % if you have one

\title{\boldmath Gravitational Waves in an $A_4$ Neutrino Mass Model}

\author{Mu-Chun Chen,}
\author{Harold J. Matias,}
\author{Cameron Moffett-Smith}
\affiliation{Department of Physics and Astronomy, 
University of California, Irvine, 
California, U.S.A}

\emailAdd{muchunc@uci.edu}
\emailAdd{hmatias@uci.edu}
\emailAdd{cmmoffet@uci.edu}

\abstract{The $A_4$ flavor symmetry has provided tremendous insight into the flavor structure of the lepton sector of the Standard Model, predicting a very good approximation to neutrino mixing angles called tri-bimaximal mixing (TBM). $A_4$ is spontaneously broken by a scalar called the flavon, and when this happens a number of degenerate vacua can form, resulting in so-called domain walls. These objects are not observed and hence need to be annihilated. This is usually done by explicitly breaking $A_4$ by adding a bias term to the scalar potential. In this paper, we construct a new model invariant under $A_4 \times \mathbb{Z}_4$ which creates cosmologically viable domain walls, lifts the degeneracy of the vacuum giving a natural mechanism for domain walls to annihilate, as well as predicts realistic neutrino mixing angles; all utilizing cross couplings between flavons. The annihilation of the domain walls, with proper choice of wall tension and the consequent bias term, leads to a gravitational wave signal that is potentially detectable in near future gravitational wave experiments, and interestingly intersects with the observed Pulsar Timing Array signal.}


\begin{document}
\maketitle
\flushbottom

\section{Introduction}

The Standard Model (SM) is one of the most successful theories of nature we have. However, we know that there are fundamental questions that it does not address. For example, whether neutrinos are Dirac or Majorana particles, the origin of neutrino masses from the discovery of their oscillations~\cite{Super-Kamiokande:1998kpq}, the mass hierarchy among generations of fermions, the large leptonic mixing compared to the quark sector, and the large number of free parameters describing the mixing. This is broadly known as the flavor puzzle. Many models based on flavor symmetries have been constructed to address the flavor puzzle.These models require invariance under $G_{SM}\times G_{F}$ where $G_F$ is a general flavor symmetry, for example, an extra $U(1)$ to explain mass hierarchies~\cite{Froggatt:1978nt}. In recent times, non-Abelian discrete symmetries~\cite{Ma:2001dn} and their modular variants~\cite{Feruglio:2017spp,Almumin:2022rml,Kobayashi:2023zzc} have gained significant interest given the observed large neutrino mixing and the predictivity afforded by these symmetries.

Flavor symmetries are spontaneously broken by the vacuum expectation value (VEV) of a scalar charged under the flavor symmetry group called the flavon. To analyze the physics of spontaneous symmetry breaking, we construct a scalar potential invariant under the flavor symmetry. Often, a $\mathbb{Z}_{2}$ symmetry is used to prevent cross couplings between flavons, as well as cubic couplings, which would further complicate the analysis. However, in the case of $A_4$, flavon cross couplings have been shown to be advantageous for generating  a nonzero $\theta_{13}$~\cite{Pascoli:2016eld}. Thus, in general, it is important to consider cross couplings between flavons.

The flavor symmetry models can be tested at neutrino experiments and rare decay searches. If the flavor symmetry scale is low, these models may also be tested at the LHC. In addition, these models can lead to interesting cosmological signatures~\cite{Ghoshal:2026hev,Gouttenoire:2025ofv,King:2023wkm, Jana:2025vyb}. When discrete symmetries are spontaneously broken, cosmological objects called domain walls will form~\cite{Zeldovich:1974uw}. A domain wall is a topological field configuration in which the vacuum is separated into regions where the VEVs of the field differ. This is caused by a degeneracy in vacuum solutions of the flavon potential. However, we want to get rid of these domain walls as they are not observed in nature. To do so, a bias term is added to lift the degeneracy of the vacuum solutions of the flavon potential. Their annihilation would produce gravitational waves which could be detected in the near future.\footnote{It is possible that the time for domain wall collapse can lead to over dense regions in which walls which are shrinking can reach their Schwarzchild radius and form primordial black holes(PBH) in a process called \enquote{catastrogenesis}~\cite{Gelmini:2022nim, Gelmini:2023ngs}. The formation of such PBH can be problematic as there may be an overproduction of them, in conflict with cosmological data. However, we are guaranteed that there won't be an overproduction of PBH based on the derived limits on the annihilation time in Sec. \ref{Section 1}. For more details please refer to~\cite{Gouttenoire:2023gbn,Gouttenoire:2025ofv,Ferreira:2024eru}. } A bias term can be induced in multiple ways, for example, higher dimensional operators originating from quantum gravity can be introduced~\cite{Jueid:2023cgp}. Additionally, anomalous symmetries are able to partially lift the degeneracy of the 
vacua~\cite{Preskill:1991kd, Chigusa:2018hhl}, or, in the $A_4$ example in particular, by modifying right handed Majorana neutrino mass term~\cite{Gelmini:2020bqg}, which also can lead to a nonzero $\theta_{13}$.  Hence we see a connection between the flavon cross couplings and the bias term which has yet to be investigated. We investigate this connection and generate the bias term in a more natural way which respects the parent $A_4$ symmetry. By adding trilinear terms and flavon cross couplings, we are able to  effectively induce a bias that respects the parent symmetry and hence provides a natural way to fully lift the vacuum degeneracy; in contrast to explicitly breaking the parent symmetry to lift the degeneracy~\cite{Gelmini:2020bqg, Jueid:2023cgp,Zeldovich:1974uw,Vilenkin:1981zs,Sikivie:1982qv,Ghoshal:2025gci,Barman:2022yos}. To the best of our knowledge,  this is the first presentation of a model symmetric under $A_4\times \mathbb{Z}_4$ to lift the vacuum degeneracy without explicitly breaking $A_4 \times \mathbb{Z}_4$ while also generating realistic leptonic mixing.

The paper is organized as follows. In Sec. \ref{Section 1}, we  provide a general review of domain walls in a $\mathbb{Z}_2$ toy model, discussing general features and properties of domain walls, including the derivation of their tension. Then, in Sec. \ref{Section 2} we briefly review flavor symmetries, demonstrating how tri-bimaximal mixing arises naturally from an $A_4$ flavor symmetry. In Sec. \ref{Section 3} we introduce the most general renormalizable flavon potential invariant under $A_4\times\mathbb{Z}_4$ with two flavon triplets, and calculate the ground states for each flavon. We then look at the effect of flavon cross couplings on the vacuum solutions in Sec. \ref{Section 4}, deriving the modified vacuum in terms of our model parameters. Next, we present our particle model and calculate the Pontecorvo-Maki-Nakagawa-Sakata (PMNS) matrix and mixing angles in Sec. \ref{Section 5}. Finally, we describe the production of gravitational waves due to the annihilation of domain walls with a bias calculated using parameters from our particle model in, and see how domain wall tension affects detectability of gravitational waves in Sec. \ref{Section 6}. In Sec. \ref{sec:conclusion} we summarize the results and main takeaways of this paper.

\section{Domain Walls} \label{Section 1}

Domain walls are two-dimensional interfaces between regions of space. They arise from field configurations which interpolate between different vacua at each region. If the universe exhibited discrete symmetries at early times, then as the universe expanded and cooled, the discrete symmetry of the potential breaks, and the vacuum expectation value of the field is forced to fall into one of the resulting degenerate minima. Domain walls, then, are the stable field configurations whose minima take different values on each region separated by the domain wall. 

The simplest example of this scenario comes from the potential for a real scalar field, $\phi$, with a Lagrangian invariant under $\mathbb{Z}_2$ given by
\begin{equation}
    \mathcal{L}= \frac{1}{2}\partial_{\mu}\phi\partial^{\mu}\phi - \frac{\lambda}{4}(\phi^2-v_{\phi}^2)^2.
    \label{eq:DWPot}
\end{equation}
This potential has two minima at $\pm v_{\phi}$. In the simple case of a planar domain wall, the field configuration would depend only on the direction perpendicular to this domain wall. We can take this direction to be the z-axis, say, in which case the field equation takes the form 
\begin{equation}
    \frac{d^2\phi}{dz^2}-\frac{dV}{d\phi}=0.
\end{equation}
A solution to this equation that gives a field configuration with a stable domain wall is given by
\begin{equation}
   \phi(z) = v_{\phi} \tanh \Bigg(\sqrt{\frac{\lambda}{2}}v_{\phi} (z-z_0)\Bigg),
\end{equation}
where $z_0$ specifies the location of the domain wall. The width of the domain wall is 
\begin{equation}
    \delta = \left(\sqrt{\frac{\lambda}{2}}v_{\phi}\right)^{-1} \; .
\end{equation}
We can calculate the stress-energy tensor $T_{\mu \nu}$, and by integrating the 00 component over the direction perpendicular to the wall, we get the surface tension of the domain wall, following the convention of~\cite{Saikawa:2017hiv} 
\begin{equation}
    \sigma = \int_{-\infty}^{\infty}dz\,T_{00}=\frac{4}{3}\sqrt{\frac{\lambda}{2}}v_{\phi}^3=f_{\sigma}\, v_{\phi}^3 \; ,
\end{equation}
where $f_{\sigma}$ is model dependent.

Now, we will introduce a bias term to lift the degeneracy of the vacua, 
\begin{equation}
    V_{b}=\varepsilon_b \,v_{\phi}^4 \; ,
    \label{eq:6}
\end{equation}
which arises from a cubic term that explicitly breaks the original $\mathbb{Z}_2$ symmetry of the toy model. Hence, the $\mathbb{Z}_2$ symmetry is an approximate symmetry if $\varepsilon_b$ is small. 
The dynamics of the domain wall are characterized by the domain wall tension $\sigma$ and the potential bias term $V_b$.

The potential in Eq.~\ref{eq:DWPot} is valid only for temperatures sufficiently below the critical temperature $T_c$ at which the phase transition occurs. The critical temperature for the potential is calculated to be $T_c =2v_{\phi}$~\cite{Gelmini:1988sf}. Fluctuations in $\phi$ become large near the critical temperature, with regions fluctuating between $\pm v_{\phi}$. Assuming a radiation dominated universe, as it cools and drops below the critical temperature, fluctuations become rare and patches of the vacuum become fixed at exponentially suppressed fluctuations. 

There are two forces governing the dynamics, the tension force and the frictional force, which we take to be negligible. The time in which the friction force is negligible and the dominant force is the tension force is called the scaling regime. During the scaling regime, the average radius of curvature $R$ is comparable to the Hubble Radius~\cite{Hiramatsu:2013qaa} and we have $R \sim H^{-1}\sim t $ and hence we can calculate the pressure due to tension as 
\begin{equation}
    p_T \approx \frac{\sigma}{R}=\frac{\sigma}{t} \; ,
\end{equation} as well as the pressure due to the energy bias\begin{equation}
    p_V\approx V_b=\varepsilon_bv_{\phi}^4 \; .
\end{equation}
The collapse of a domain wall occurs when the pressure due to the potential bias is comparable or larger than the pressure due to tension $p_T \approx p_V$~\cite{Saikawa:2017hiv}, hence we can calculate the annihilation time\begin{equation}
    t_{ann}=\frac{\sigma}{\varepsilon_bv_{\phi}^4}=\frac{\sigma}{V_b} \; .
\end{equation}
We can also calculate the time at which the energy density stored in the domain walls is larger than the energy density of the universe, to get an upper and lower bound on the bias parameter $\varepsilon_b$. This has been done previously~\cite{Gelmini:2020bqg}, so we quote the result: \begin{equation} \label{eq:10}
    10^{-25}\left ( \frac{\text{TeV}}{v_{\phi}}\right)<\frac{\varepsilon_b}{f_\sigma}<10^{-15}
\left(\frac{v_{\phi}}{\text{TeV}}\right) \; .
\end{equation}

\section{Flavor Symmetries} \label{Section 2}

Understanding the origin of the flavor structure of the SM is a major focus in particle physics today. Considering the success of symmetry principles in describing gauge interactions in the SM, we consider flavor symmetries as a possible origin of the flavor structure in the SM. 

Although the flavor sector does not appear to have any exact symmetries, it nevertheless contains some structure. Most obviously, fermion masses are hierarchical in both the quark and lepton sectors. Also, quark mixing is very small, which makes the Cabibbo-Kobayashi-Maskawa (CKM) matrix almost diagonal. In the lepton sector, on the other hand, mixing is very large, leading to a PMNS matrix with entries with values of roughly the same order. This motivates non-Abelian discrete symmetries as a possible origin of the observed flavor structure in the SM. 

The structure of the PMNS matrix itself offers a guide towards a resolution to the flavor puzzle. This matrix results from the mismatch between the mass and flavor eigenstates of leptons. The best fit values of the reactor and atmospheric mixing angles in degrees and their corresponding $1\sigma$ ranges are $\theta_{13}=8.56^{+0.11} _{-0.11}$ and $\theta_{23}=43.3^{+1.0} _{-0.8}$ ~\cite{Esteban:2024eli}. 
Motivated by these values, we can take as a leading order approximation the values $\theta_{13}=0^\circ$ and $\theta_{23}=45^\circ$. The most general form of the PMNS matrix satisfying this approximation (assuming a diagonal charged-lepton mass matrix) is given by 
\begin{equation}
    U=\begin{pmatrix}
        c_{12} & s_{12} & 0 \\
        -\frac{s_{12}}{\sqrt{2}} & \frac{c_{12}}{\sqrt{2}} & \frac{1}{\sqrt{2}} \\
        -\frac{s_{12}}{\sqrt{2}} & \frac{c_{12}}{\sqrt{2}} & -\frac{1}{\sqrt{2}}
    \end{pmatrix} \; .
\end{equation}
where $c_{12}=\cos{\theta_{12}}$ and $s_{12}=\sin{\theta_{12}}$. Incidentally, this corresponds to a bimaximal mixing between $\nu_\mu$ and $\nu_\tau$. That is,
\begin{equation}
    U_{e3}=0, \quad |U_{\mu 3}|= |U_{\tau 3}|= \frac{1}{\sqrt{2}} \implies \nu_3 = \frac{1}{\sqrt{2}}(\nu_\mu-\nu_\tau) \; .
\end{equation}

The solar angle $\theta_{12}=33.68^{+0.73} _{-0.70}$ ~\cite{Esteban:2024eli} also within 1$\sigma$ is very close to what we get when we assume a trimaximal mixing between $\nu_2$ and all neutrino flavors. From the mixing
\begin{equation}
    \nu_2 = \frac{1}{\sqrt{3}}(\nu_e + \nu_\mu + \nu_\tau) \implies |U_{e2}|= |U_{\mu2}|= |U_{\tau2}|= \frac{1}{\sqrt{3}},
\end{equation}
we get a solar angle $\theta_{12}= \sin ^{-1}(\frac{1}{\sqrt{3}})=35.3^\circ$, which is within the bounds of the observed value. We finally end up with the TBM form of the lepton mixing matrix
\begin{equation}
    U_{TBM}=\begin{pmatrix}
        \frac{2}{\sqrt{6}} & \frac{1}{\sqrt{3}} & 0 \\
        -\frac{1}{\sqrt{6}} & \frac{1}{\sqrt{3}} & \frac{1}{\sqrt{2}} \\
        -\frac{1}{\sqrt{6}} & \frac{1}{\sqrt{3}} & -\frac{1}{\sqrt{2}}
    \end{pmatrix}.
\end{equation} 

\begin{table}[ht]
\centering
\begin{tabular}{cccccccccc}
\hline
Fields & $\xi$ & $\phi$ & $\chi$ & L & N & H & $e_R$ & $\mu_R$ & $\tau_R$ \\
\hline
$A_4$ & $\textbf{1}$ & $\textbf{3}$ & $\textbf{3}$ & $ \textbf{3}$ & $\textbf{3}$ & $\textbf{1}$ & $\textbf{1}$ & $\textbf{1}^{\prime \prime}$ & $\textbf{1}^{\prime }$ \\
$G_{\text{SM}}$ & $(\textbf{1},\textbf{1})_0$ & $(\textbf{1},\textbf{1})_0$ & $(\textbf{1},\textbf{1})_0$ & $(\textbf{1},\textbf{2})_{\nicefrac{-1}{2}}$ & $(\textbf{1},\textbf{1})_0$ & $(\textbf{1},\textbf{2})_{\nicefrac{1}{2}}$ & $(\textbf{1},\textbf{1})_{-1}$ & $(\textbf{1},\textbf{1})_{-1}$ & $(\textbf{1},\textbf{1})_{-1}$ \\
\hline
\end{tabular}
\caption{Particle content and representation assignments in an $A_4$ model with type-1 seesaw. $G_\text{SM}= \text{SU}(3)_C \times \text{SU}(2)_L \times \text{U}(1)_Y$.}
\label{tab:A4particlecontent}
\end{table}

To see how TBM naturally arises from $A_4$, we consider the Lagrangian
\begin{align}
    \mathcal{L} \supset y_D \left(\bar{L}N\right)_{\textbf{1}}\tilde{H} + \frac{1}{2} x_c \left( \overline{N^c} N \right)_{\textbf{1}}\xi +  \frac{1}{2} x_b \left( \overline{N^c} N \right)_{\textbf{3}} \chi \\ \notag+
    \frac{y_e}{\Lambda}\left( \phi \bar{L}\right)_{\textbf{1}}e_R H\;+\frac{y_{\mu}}{\Lambda}\left( \phi \bar{L}\right)_{\textbf{1}^{\prime}}\mu_R H+\frac{y_{\tau}}{\Lambda}\left( \phi \bar{L}\right)_{\textbf{1}^{\prime\prime}}\tau_R H\;+h.c.
\end{align}
In the Altarelli-Feruglio (AF) basis, the flavons $\phi$, $\chi$, and $\xi$ acquire the VEVs
\begin{equation}
\langle \phi \rangle
=
v_\phi
\begin{pmatrix}
1 \\[2pt]
0 \\[2pt]
0
\end{pmatrix},
\qquad
\langle \chi \rangle
=
\frac{v_\chi}{\sqrt{3}}
\begin{pmatrix}
1 \\[2pt]
1 \\[2pt]
1
\end{pmatrix},
\qquad
\langle \xi \rangle
=
u \; .
\end{equation} 
After EWSB and applying the $A_4$ tensor product contractions in the AF basis, we get a diagonal charged lepton mass matrix given by
\begin{equation}
M_\ell
=
\frac{v\,v_\phi}{\Lambda}\,
\mathrm{diag}\!\left(y_e,\,y_\mu,\,y_\tau\right),
\end{equation}
up to unphysical phase redefinitions of the lepton fields. This implies that the unitary transformation diagonalizing the charged leptons is just the identity. So the PMNS matrix becomes
\begin{equation}
    U_{\text{PMNS}}=U_l^\dagger U_\nu = U_\nu\;.
\end{equation}

After integrating out the right-handed heavy neutrinos, we get an effective light Majorana neutrino mass term
\begin{equation}
\mathcal{L}_\nu \;\supset\;
\frac{1}{2}\,\overline{\nu_L^{c}}\,
M_\nu\,\nu_L
\;+\;\text{h.c.},
\end{equation}
where
\begin{equation}
M_\nu
= -M_D M_N^{-1} M_D^T \;.
\end{equation}
Taking the $A_4$ tensor products in the AF basis, the Dirac neutrino mass takes the form
\begin{equation}
M_D= \frac{v}{\sqrt{2}}
    \begin{pmatrix}
1 & 0 & 0 \\[4pt]
0 & 0 & 1 \\[4pt]
0 & 1 & 0
\end{pmatrix} =
\frac{v}{\sqrt{2}} P_{\text{23}}\;,
\end{equation} 
where we express the matrix above as $P_{23}$ since it permutes the second and third flavor eigenstates. The heavy right-handed Majorana neutrino mass matrix
may be written (up to an overall normalization) as
\begin{equation}
M_N
=
\begin{pmatrix}
c+2b & -b & -b \\[4pt]
-b & 2b & c-b \\[4pt]
-b & c-b & 2b
\end{pmatrix},
\qquad
c = \frac{x_c\,u}{2}\,,
\qquad
b = \frac{x_b\,v_\chi}{4\sqrt{3}}\,.
\end{equation}

To diagonalize this matrix, we find its eigenvectors. Placing the eigenvectors as columns, we end up with the tri-bimaximal mixing matrix
\begin{equation}
U_{TBM}=\begin{pmatrix}
        \frac{2}{\sqrt{6}} & \frac{1}{\sqrt{3}} & 0 \\
        -\frac{1}{\sqrt{6}} & \frac{1}{\sqrt{3}} & \frac{1}{\sqrt{2}} \\
        -\frac{1}{\sqrt{6}} & \frac{1}{\sqrt{3}} & -\frac{1}{\sqrt{2}}
    \end{pmatrix}\;,
\end{equation}
which satisfies
\begin{equation}
U_{\mathrm{TBM}}^{T}\,M_N\,U_{\mathrm{TBM}}
=
\mathrm{diag}(m_1,m_2,m_3),
\end{equation}
where the diagonal entries are the eigenvalues of $M_N$ which are given by
\begin{equation}
m_1 = c+3b,
\qquad
m_2 = c,
\qquad
m_3 = -c+3b .
\end{equation}
We know that the eigenvectors of a matrix and its inverse are the same. Also, their eigenvalues simply get rescaled to the value of their inverses. Thus, $M_N$ and $M_N^{-1}$ are diagonalized by the same matrix $U_{TBM}$ but with the inverse eigenvalues, e.g., 
\begin{equation}
    U_{TBM}^T M_N^{-1} U_{TBM} 
    = \text{diag}(m_1^{-1},m_2^{-1},m_3^{-1})\;.
\end{equation}
Furthermore, from the properties of $M_D = \frac{v}{\sqrt{2}} P_{\text{23}}$ and $M_N^{-1}$, we can show that $M_\nu=-M_D M_N^{-1} M_D^T$ is also diagonalized by $U_{TBM}$. Thus, in the basis where the charged-lepton mass matrix is diagonal, we have
\begin{equation}
    U_{PMNS} = U_l^\dagger U_\nu = U_\nu \simeq U_{TBM}\;,
\end{equation}
and 
\begin{equation}
    U_{TBM}^T M_\nu U_{TBM} 
    = \frac{v^2}{2} \text{diag}(m_1^{-1},m_2^{-1},m_3^{-1})\;.
\end{equation}

The TBM form of the PMNS matrix was ruled out by experiment, especially by the observation of a non-zero reactor angle $\theta_{13}$. The approximate TBM mixing in the lepton sector nevertheless motivated the introduction of non-Abelian discrete symmetries. In particular, TBM implies a residual $\mathbb{Z}_2$ symmetry in the neutrino sector and a $\mathbb{Z}_3$ symmetry in the charged-lepton sector. The smallest non-Abelian discrete group that contains both $\mathbb{Z}_2$ and $\mathbb{Z}_3$ as subgroups is $A_4$. So even though tri-bimaximal mixing is ruled out by experiment, it still works as a first-order approximation to the PMNS matrix and hence non-Abelian discrete symmetries still play a major role in potentially accounting for leptonic mixing.

In our model, we show that by including all renormalizable terms in the flavon potential, including the cubic and cross couplings, we can get realistic mixing angles. The same set of parameters also provides a mechanism for domain wall annihilation, leading to observable gravitational wave signals.

\section{Flavon Sector} \label{Section 3}
\subsection{Flavon Potential}

The symmetry $A_4$ contains three singlet representations, the trivial singlet \textbf{1}, two nontrivial singlets $\textbf{1}^\prime$, $\textbf{1}^{\prime \prime}$, and one  triplet representation \textbf{3}. Being the lowest order group with a three-dimensional irreducible representation, first studied in~\cite{Ma:2001dn}, it has attracted a lot of attention due to its predictions for leptonic mixing~\cite{Altarelli:2005yx,King:2013eh, Feruglio:2008ht, Heinrich:2018nip, Chen:2012st}. Even though $A_4$ predicts TBM, it is still worth looking into for theoretical reasons, not only as a building block towards modular flavor symmetries~\cite{Feruglio:2017spp}, but also proves to hold richer phenomenology in the form of cross couplings~\cite{Pascoli:2016eld,Pascoli:2016wlt}, as will be discussed in Sec. \ref{Section 5}.  The tensor products of the representations of $A_4$ are given in Appendix \ref{appendixA}. 

We build a model that has $A_4\times\mathbb{Z}_4$ symmetry. We assign the flavon, $\phi$, to a  triplet representation of $A_4$, which carries charge $+1$ under $\mathbb{Z}_4$. We label the three components of $\phi$ as $\phi=(\phi_1,\phi_2,\phi_3)$ and write the most general renormalizable flavon potential invariant under $A_4\times\mathbb{Z}_4$, 
\begin{eqnarray}
    V(\phi)& = & \frac{1}{2}\mu_{\phi}^2(\phi\phi)_{\textbf{1}}\;+ \frac{1}{4}
\left[ f_1(\phi\phi)_{\mathbf{1}}^2\:+f_2(\phi\phi)_{\mathbf{1}^\prime}(\phi\phi)_{\textbf{1}^{\prime \prime}}\;+f_3 \left((\phi\phi)_{\textbf{3}_S}(\phi\phi)_{\textbf{3}_S} \right)_{\textbf{1}}\right]\; \nonumber \\
& & +\frac{f_4}{3}((\phi\phi)_{\textbf{3}_S}\phi)_{\textbf{1}} \; ,
\end{eqnarray}
 where $f_1$, $f_2$, $f_3$, and $f_4$ are free couplings, and $\mu_{\phi}^{2}$ is a mass-like term. Expanding the products in the Ma-Rajasekaran (MR) basis, we get the following potential, \begin{eqnarray}
    V(\phi)& = & \frac{1}{2}\mu_{\phi}^2
(\phi_1^2+\phi_2^2+\phi_3^2)\;+\frac{f_1+f_2}{4}(\phi_1^2+\phi_2^2+\phi_3^2)^2\;+\frac{3(f_3-f_2)}{4}(\phi_1^2\phi_2^2+\phi_1^2\phi_3^2+\phi_2^2\phi_3^2)\; \nonumber\\
& & +\sqrt{3}f_4\phi_1\phi_2\phi_3 \; ,
\end{eqnarray}
which interestingly corresponds to the fully renormalizable potential invariant under $S_4$ in which domain walls have been recently investigated~\cite{Fu:2024jhu}.Using the following relations, 
\begin{align}
I_1=\phi_1^2+\phi_2^2+\phi_3^2 \; , \quad I_2=\phi_1^2\phi_2^2+\phi_1^2\phi_3^2+\phi_2^2\phi_3^2 \; , \nonumber\\[1em]
g_1=f_1+f_2 \; , \quad 
g_2=\frac{3(f_3-f_2)}{2} \; , \quad g_3=\sqrt{3}f_4 \; . \nonumber
\end{align}
we can express the potential as
\begin{align}
    V(\phi)=\frac{1}{2}\mu_{\phi}^2 I_1+\frac{g_1}{4}I_1^2+\frac{g_2}{2}I_2+g_3\phi_1\phi_2\phi_3 \; .
    \label{equation18}
\end{align}
We write the potential for $\chi$ in the exact same form with $(g_{i}, \mu_{\phi}^{2})$ being replaced by $(\tilde{g}_{i}, \mu_{\chi}^{2})$, and the absence of a cubic term.

\begin{align}
    V(\chi)=\frac{1}{2}\mu_{\chi}^2
    \tilde{I}_1+\frac{\tilde{g_1}}{4}\tilde{I}_1^2+\frac{\tilde{g_2}}{2}\tilde{I}_2 \; ,
    \label{equation19}
\end{align} 
where $\tilde{I}_1$ and $\tilde{I}_2$ have the same functional forms as $I_{1}$ and $I_{2}$ except with $\phi$ being replaced by $\chi$. We are focusing on a model with zero temperature. Thermal perturbations will have an effect on the potential, for example the universe finding itself in a metastable state, but as the universe cools, it tunnels to the true degenerate vacuum states through the nucleation of bubbles~\cite{Paul:2020wbz}. However the two scenarios are decoupled and the formation of bubbles won't affect the spectrum of gravitational waves from the collapse of domain walls. We are interested in the properties of the flavon after thermal interactions become negligible. However if the bubbles are formed, the signal will be distinct from the domain wall signal due to the frequency range in which the signal peaks.

Next, we minimize the potential. This is non-trivial due to the presence of the cubic term in $V(\phi)$. To circumvent this, we search for vacuum solutions that respect the residual $\mathbb{Z}_2$ and $\mathbb{Z}_3$ symmetries after $A_4$ is spontaneously broken. That is we look for solutions to $G_i\langle\phi\rangle=\langle\phi\rangle$ where $G_i$ are generators for residual $\mathbb{Z}_2$ and $\mathbb{Z}_3$ groups. The residual $\mathbb{Z}_2$ and $\mathbb{Z}_3$ symmetries, correspond  to the neutrino sector and to the charged lepton sector respectively~\cite{Fu:2024jhu}. We find in MR basis, 
\begin{align}
    \left \{ \begin{pmatrix}
    1\\0\\0
    \end{pmatrix},\begin{pmatrix}
    0\\1\\0
    \end{pmatrix},\begin{pmatrix}
    0\\0\\1
    \end{pmatrix},\begin{pmatrix}
    -1\\0\\0
    \end{pmatrix},\begin{pmatrix}
    0\\-1\\0
    \end{pmatrix},\begin{pmatrix}
    0\\0\\-1
    \end{pmatrix}
    \right \}v_{\chi},
\end{align}
 \begin{align}
    \left \{ \begin{pmatrix}
    1\\1\\1
    \end{pmatrix},\begin{pmatrix}
    1\\-1\\-1
    \end{pmatrix},\begin{pmatrix}
    -1\\1\\-1
    \end{pmatrix},\begin{pmatrix}
    -1\\-1\\1
    \end{pmatrix}
    \right \}v_{\phi -} \; ,
\end{align}
and
\begin{align}
    \left \{\begin{pmatrix}
    -1\\1\\1
    \end{pmatrix},\begin{pmatrix}
    1\\-1\\1
    \end{pmatrix},\begin{pmatrix}
    1\\1\\-1
    \end{pmatrix},\begin{pmatrix}
    -1\\-1\\-1
    \end{pmatrix}\right \}v_{\phi +}
\end{align}
being a set of $\mathbb{Z}_2$ preserving solution and two sets of $\mathbb{Z}_3$ preserving solutions respectively. $v_{\phi \pm}$ and $v_{\chi}$ are determined by plugging each solution into the potential and minimizing with respect to the VEV($v_{\phi \pm}$ or $v_{\chi}$), to have the form \begin{align}
    v_{\chi}=\frac{\mu_{\chi}}{\sqrt{\tilde{g_1}}} \; ,
\end{align}
and \begin{align}
    v_{\phi \pm}=\frac{\mu_{\phi}}{\sqrt{3g_1+2g_2}}\left ( \sqrt{1+a^2}\pm a\right) \; ,
\end{align}
where $a=\frac{g_3}{2\mu_{\phi}\sqrt{3g_1+2g_2}}$, which we see is in agreement with current discussions on cubic couplings~\cite{Fu:2024jhu}. We have calculated these solutions in the Altarelli-Feruglio basis in Appendix \ref{appendix2}. Note, one can enforce, through the signs of the coupling constants, whether the alignment falls into one set of $\mathbb{Z}_3$ solutions or the other set by analyzing the eigenvalues of the Hessian matrix and enforcing stability of one solution over another as given by \begin{equation}
    (M^2_{\phi})_{ij} = \frac{\partial^2V(\phi)}{\partial \phi_{i}\partial\phi_j}\bigg|_{\langle \phi \rangle} \; ,
\end{equation}
which has eigenvalues $m_1^2,\,m_2^2,\,m_3^2$. Upon plugging in our potentials and diagonalizing we get \begin{equation}
        m_{\chi_1}^2=2(\tilde{f_1}+\tilde{f_2})v_{\chi^2} \; , \quad m_{\chi_2}^2=m_{\chi_3}^2=(\tilde{f_2}-\tilde{f_3})v_{\chi^2} \; ,
\end{equation}
and
\begin{equation}
    \begin{aligned}
      &  m_{{\phi_1}_{\pm}}^2=2\mu_{\phi}^2(1+a^2\pm a\sqrt{1+a^2}) \; , \\[1em] & m_{{\phi_2}_{\pm}}^2=m_{{\phi_3}_{\pm}}^2=\frac{f_3-f_2}{2f_2+f_1+f_3}\mu_{\phi}^2\left[ 1-2a(a\pm \sqrt{1+a^2}) \left ( 1+2\frac{f_1+f_2}{f_2-f_3}\right )\right] \; ,
    \end{aligned}
\end{equation}
agreeing with previous literature~\cite{Fu:2025qhf,Fu:2024jhu}.

\subsection{Flavon Cross Couplings} \label{Section 4}
Including operators that mix flavons has been demonstrated to have interesting phenomenological consequences, and therefore must be taken into consideration~\cite{Pascoli:2016eld, Pascoli:2016wlt, deMedeirosVarzielas:2025byb}. For what follows, we will work in the Altarelli-Feruglio basis. We assume a flavor symmetry of $A_4$ augmented with $\mathbb{Z}_4$ under which only $\chi$ is negatively charged. We then write the full renormalizable potential that couples both $\phi$ and $\chi$, with all terms respecting $A_4\times \mathbb{Z}_4$,
\begin{equation}
\begin{aligned}
    V(\phi,\chi)=\frac{1}{2}\varepsilon_1(\chi\chi)_{\textbf{1}}(\phi\phi)_{\textbf{1}}+\frac{1}{4}\varepsilon_2(\chi\chi)_{\textbf{1}^{\prime}}(\phi\phi)_{\textbf{1}^{\prime\prime}}+ \frac{1}{4}\varepsilon_2^*(\chi\chi)_{\textbf{1}^{\prime\prime}}(\phi\phi)_{\textbf{1}^{\prime}}
    \\+ \frac{1}{2}\varepsilon_3\left [(\chi\chi)_{\textbf{3}}(\phi\phi)_{\textbf{3}} \right ]_{\textbf{1}} + \frac{1}{3}\varepsilon_4[(\chi\chi)_{\textbf{3}}\phi]_{\textbf{1}} \; ,
\end{aligned}
\end{equation}
where all triplet contractions are assumed to be symmetric, and we assume the coupling constants to be small enough to preserve the residual $\mathbb{Z}_2$ and $\mathbb{Z}_3$ symmetries at leading order.
\par In principle, we can go to higher order terms of dimension 5, 6, or 7; but we conclude that the effects of these higher order terms will be negligible. This is because the cross couplings are assumed to be of $\mathcal{O}(0.1)$ so that we can achieve such order corrections to the TBM matrix~\cite{Pascoli:2016eld}. Further higher dimensional operators will be suppressed by subsequent powers of $\Lambda$ which is taken to be a very high energy cutoff, such that each of the higher dimensional operators will be suppressed by such a factor yielding negligible corrections to the flavor sector.

We choose the following alignments to expand around in the Altarelli-Feruglio basis, \begin{equation}
    \langle\chi\rangle=\begin{pmatrix}
    1\\1\\1
\end{pmatrix}\frac{v_{\chi}}{\sqrt{3}} \; , 
\quad\langle\phi\rangle=\begin{pmatrix}
    1\\0\\0
\end{pmatrix}\;v_{\phi} \; ,
\end{equation} 
assuming the perturbations to the vacua are small, we get the form, 
\begin{equation}
    \langle\chi\rangle=\begin{pmatrix}
    \frac{v_{\chi}}{\sqrt{3}}+\delta v_{\chi_1}
    \\\frac{v_{\chi}}{\sqrt{3}}+\delta v_{\chi_2}
    \\\frac{v_{\chi}}{\sqrt{3}}+\delta v^*_{\chi_2}
\end{pmatrix}\; , 
\quad\langle\phi\rangle=\begin{pmatrix}
    v_{\phi}+\delta v_{\phi_1}
    \\\delta v_{\phi_2}
    \\\delta v^*_{\phi_2}
\end{pmatrix} \; .
\end{equation}
We can plug the perturbed vacua into their respective potentials as well as the potential coupling $\phi$ and $\chi$. We keep terms quadratic in the vacuum shifts for the separate potentials $V(\phi)$ and $V(\chi)$ because the first order shifts are trivial and do not affect the vacuum structure. For the cross coupling potential $V(\phi,\chi)$, we go to fist order shifts as higher order shifts are negligible since the coupling constants are taken to be small. Calculations of this sort have been done in detail~\cite{Pascoli:2016eld, Pascoli:2016wlt}, however we consider a cubic term that exists when only one of the flavons are non-trivially charged under $\mathbb{Z}_4$. Thus, our methodology is practically equivalent, and hence we will display results. We find that the shifts in the vacuum caused by cross couplings are given by the following expressions
\begin{equation}
\begin{aligned}
&
\delta v_{\phi_1}=0,\quad \delta v_{\phi_2}=\frac{3v_{\chi}^2\varepsilon_2}{2\sqrt{3}f_4-6(f_2+f_3+2(f_1+a(a^2+\sqrt{1+a^2})f_1+a(a+\sqrt{1+a^2})f_3))v_{\phi_-}} \;, \\[1em]
&    \delta v_{\phi_2}^* = \frac{3v_{\chi}^2\varepsilon_2^*}{2\sqrt{3}f_4-6(f_2+f_3+2(f_1+a(a^2+\sqrt{1+a^2})f_1+a(a+\sqrt{1+a^2})f_3))v_{\phi_-}}
   \; , \\[1em]
&    \delta v_{\chi_1}=\frac{-2v_{\phi_-}(3\sqrt{3}v_{\phi_-}\varepsilon_3+2\varepsilon_4)}{3(4\tilde{f_1}+\tilde{f_2}+3\tilde{f_3})v_\chi}, \quad \delta v_{\chi_2}=\delta v_{\chi_2}^*=\frac{v_\phi(3\sqrt{3}v_{\phi_-}\varepsilon_3+2\varepsilon_4)}{3(4\tilde{f_1}+\tilde{f_2}+3\tilde{f_3})v_\chi}
\; ,
\end{aligned}
\end{equation}
which written in a cleaner way is
\begin{equation}
    \begin{aligned}
        \delta v_{\phi1}=0 \; ,\quad \delta v_{\phi2}=v_{\phi_-}\varepsilon_{\phi} \; ,\quad \delta v_{\phi2}^*=v_{\phi_-}\varepsilon_{\phi}^* \; ,\\
        \delta v_{\chi1}=-2v_{\chi}\varepsilon_{\chi} \; ,\quad \delta v_{\chi2}=\delta v_{\chi2}^*=v_{\chi}\varepsilon_{\chi} \; ,
    \end{aligned}
\end{equation}
where 
\begin{align}
    \varepsilon_{\phi}=\frac{3v_{\chi}^2\varepsilon_2}{(2\sqrt{3}f_4-6(f_2+f_3+2(f_1+a(a^2+\sqrt{1+a^2})f_1+a(a+\sqrt{1+a^2})f_3))v_{\phi_-})v_{\phi_-}}
    \label{eq:19}
\end{align}is in general a complex parameter parametrized by $\varepsilon_{\phi}=|\varepsilon_{\phi}|e^{i\theta_{\phi}}$, and 
\begin{align}
    \varepsilon_{\chi}=\frac{v_{\phi_-}(3\sqrt{3}v_{\phi_-}\varepsilon_3+2\varepsilon_4)}{3(4\tilde{f_1}+\tilde{f_2}+3\tilde{f_3})v_\chi^2} \label{eq:20}
\end{align}is a real parameter. An interesting feature is that the bias between $\mathbb{Z}_2$ domain walls is completely determined by $\varepsilon_4$ which can be demonstrated in the limiting case that $\varepsilon_4$ goes to zero, in which case there is no bias.

\section{Particle Model and Neutrino Parameters} \label{Section 5}
\subsection{Particle Model}
In our model invariant under $A_4\times \mathbb{Z}_4$, the left handed lepton doublets $L$, and three right hand neutrinos $ N=\begin{pmatrix}
    N_1,N_2,N_3
\end{pmatrix}$ transform as triplets of $A_4$. The $SU(2)_L$ Higgs doublet transforms as a trivial singlet $\tilde{H}\sim\textbf{1}$, and the right handed leptons as singlets $e_R\sim\textbf{1},\: \mu_R\sim \textbf{1}'',\: \tau_R \sim\textbf{1}'$. 
The two triplet flavons, $\phi$ and $\chi$,  assume the VEV alignments 
$\langle\phi\rangle= (1+\varepsilon_{\phi},\:\varepsilon_{\phi},\: \varepsilon_{\phi}^*)v_{\phi},\ \langle\chi\rangle= (1-2\varepsilon_{\chi},\:1+\varepsilon_{\chi},\:1+\varepsilon_{\chi})\frac{v_{\chi}}{\sqrt{3}}$, as shown in Sec. \ref{Section 4}. Here we assume  $v_{\phi}=v_{\phi_-}$ for ease of notation. We summarize the particle content in Table~\ref{tab:particlecontent}.

\begin{table}[ht]
\centering
\begin{tabular}{cccccccccc}
\hline
Fields & $\eta$ & $\phi$ & $\chi$ & L & N & H & $e_R$ & $\mu_R$ & $\tau_R$ \\
\hline
$A_4$ & $\textbf{1}$ & $\textbf{3}$ & $\textbf{3}$ & $ \textbf{3}$ & $\textbf{3}$ & $\textbf{1}$ & $\textbf{1}$ & $\textbf{1}^{\prime \prime}$ & $\textbf{1}^{\prime }$ \\
$\mathbb{Z}_4$ & -1 & 1 & -1 & $i$ & $i$ & 1 & $i$ & $i$ & $i$ \\
\hline
\end{tabular}
\caption{Particle Content and Representation Assignments}
\label{tab:particlecontent}
\end{table}

We can then write the Lagrangian \begin{align}
    \mathcal{L} \supset y_D \left ( \bar{L}N\right)_{\textbf{1}}\tilde{H}\;+y_N\left[ \left( \bar{N}N^c\right)_{\textbf{3}}\chi\right]_{\textbf{1}}\; +\frac{1}{2}y_1 \eta \bar{N}^cN \notag \\+
    \frac{y_e}{\Lambda}\left( \phi \bar{L}\right)_{\textbf{1}}e_R H\;+\frac{y_{\mu}}{\Lambda}\left( \phi \bar{L}\right)_{\textbf{1}^{\prime}}\mu_R H+\frac{y_{\tau}}{\Lambda}\left( \phi \bar{L}\right)_{\textbf{1}^{\prime\prime}}\tau_R H\;+h.c. \; ,
\end{align}
where $\Lambda$ is a high energy scale and $\eta$ is a singlet scalar with $\langle\eta\rangle=v_{\eta}$. 
\par
Expanding about the perturbed vev of $\phi$, upon electroweak symmetry breaking, we get the charged lepton mass matrix,
\begin{equation}
   M_{\ell}= \frac{v_H v_{\phi}}{\sqrt{2}\Lambda}\begin{pmatrix}
y_e&\varepsilon_{\phi}y_{\mu}&\varepsilon_{\phi}^*y_{\tau}\\\varepsilon_{\phi}^*y_e&y_{\mu}&\varepsilon_{\phi}y_{\tau}\\\varepsilon_{\phi}y_e&\varepsilon_{\phi}^*y_{\mu}&y_{\tau}
    \end{pmatrix} \; ,
\end{equation}
where $v_H$ is the vev of the Higgs. 
Similarly, we derive Dirac and Majorana mass matrices,
\begin{equation}
    M_D=\frac{y_Dv_H}{2}\begin{pmatrix}
        1&0&0\\0&0&1\\0&1&0
    \end{pmatrix} \; ,
\end{equation} and
\begin{equation}
\begin{aligned}
&   M_N= \frac{1}{2\sqrt{3}} \times 
    \\
   & \begin{pmatrix}
        y_N(1-2\varepsilon_{\chi})v_{\chi}+\sqrt{3}y_1v_{\eta}&-\frac{1}{2}y_N(1+\varepsilon_{\chi})v_{\chi}& -\frac{1}{2}y_N(1+\varepsilon_{\chi})v_{\chi}\\ -\frac{1}{2}y_N(1+\varepsilon_{\chi})v_{\chi}& \frac{1}{2}y_N(1+\varepsilon_{\chi})v_{\chi}&
        -y_N(1-2\varepsilon_{\chi})v_{\chi}+\sqrt{3}y_1v_{\eta}\\-\frac{1}{2}y_1(1+\varepsilon_{\chi})v_{\chi}& 
        -y_N(1-2\varepsilon_{\chi})v_{\chi}+\sqrt{3}y_1v_{\eta}& \frac{1}{2}y_N(1+\varepsilon_{\chi})v_{\chi} \; ,       
    \end{pmatrix} \; , \\[1em]
\end{aligned}
\end{equation}
respectively. Note that this is analogous to what has been previously done, with the exception of our different definitions of our perturbations. Details can be found in~\cite{Pascoli:2016eld}.

Next, we diagonalize the charged lepton and the effective light neutrino mass matrices. We  parametrize the PMNS matrix as 
\begin{equation}
    U_{PMNS}\approx U_{\ell}^{\dagger}U_{TBM}U_{\nu}P_{\nu} \; ,
\end{equation}
where $U_{\ell}^{\dagger}$ is the correction we get from diagonalizing the charged lepton mass matrix, $U_{TBM}$ is the usual tri-bimaximal mixing matrix, $U_{\nu}$ is the correction we get from diagonalizing the 
left-handed neutrino matrix, and $P_{\nu}$ is a diagonal matrix containing the Majorana phases~\cite{Pascoli:2016wlt}. We get this form since, in our basis(Altarelli-Feruglio), $M_{\ell}$ is approximately diagonal and $M_{\nu}$ is approximately diagonalized by TBM. We find by perturbatively diagonalizing the mass matrices that
\begin{equation}
    U_{\ell}^{\dagger}\approx\begin{pmatrix}
        1&-\varepsilon_{\phi}&-\varepsilon_{\phi}^*\\
        \varepsilon_{\phi}^*&1&-\varepsilon_{\phi}\\
        \varepsilon_{\phi}&\varepsilon_{\phi}^*&1
    \end{pmatrix} \; ,\quad 
    U_{\nu}\approx\begin{pmatrix}
        1&\sqrt{2}\varepsilon_{\chi}&0\\
        -\sqrt{2}\varepsilon_{\chi}&1&0\\0&0&1
    \end{pmatrix} \; .
\end{equation}
Thus, \begin{equation}
    U_{PMNS}\approx \begin{pmatrix}
        1&-\varepsilon_{\phi}&-\varepsilon_{\phi}^*\\\varepsilon_{\phi}^*&1&-\varepsilon_{\phi}\\\varepsilon_{\phi}&\varepsilon_{\phi}^*&1
    \end{pmatrix}
    \begin{pmatrix}
        \frac{2}{\sqrt{6}}&\frac{1}{\sqrt{3}}&0\\-\frac{1}{\sqrt{6}}&\frac{1}{\sqrt{3}}&-\frac{1}{\sqrt{2}}\\-\frac{1}{\sqrt{6}}&\frac{1}{\sqrt{3}}&\frac{1}{\sqrt{2}}
    \end{pmatrix}
    \begin{pmatrix}
        1&\sqrt{2}\varepsilon_{\chi}&0\\
        -\sqrt{2}\varepsilon_{\chi}&1&0\\0&0&1
    \end{pmatrix}P_{\nu} \; .
\end{equation}
Given this, we have the mixing angles to first order in $\varepsilon_\phi$ and $\varepsilon_\chi$ to be
\begin{equation} \label{angles}
    \begin{aligned}
        \sin\,\theta_{13}\approx\sqrt{2}|\varepsilon_{\phi}\sin\,\theta_{\phi}| \; , \\
        \sin\, \theta_{12}\approx\frac{1}{\sqrt{3}}(1+2\varepsilon_\chi-2|\varepsilon_\phi|\cos\, \theta_\phi) \; , \\
        \sin \, \theta_{23}\approx\frac{1}{\sqrt{2}}(1+|\varepsilon_\phi|\cos\,\theta_\phi) \; ,
    \end{aligned}
\end{equation}
which gives TBM in the limit that $\varepsilon_\chi$ and $\varepsilon_\phi$ go to 0 as expected. Notice we can derive a sum rule as
\begin{equation}\label{sumrule}
    \sin\,\theta_{12}\approx \frac{1}{\sqrt{3}}(2\varepsilon_{\chi}-2\sqrt{2}\sin\,\theta_{23}+3)
\end{equation}It is interesting to note that a nonzero reactor angle requires $\varepsilon_\phi$ to be complex. One can also calculate the Dirac CP violating phase $\delta_{CP}$. Interestingly, this is an effect that shows up at second order, so one must be careful in this calculation. At first order, we predict maximal CP violation of $\delta_{CP}=\frac{3\pi}{2}$, we then have a shift due to higher order corrections of 
\begin{equation}
    \delta_{CP} = \frac{3\pi}{2}-2|\varepsilon_{\phi}|\,\sin\,\theta_{\phi} 
    \; .
\end{equation} 
This demonstrates that for more detailed and accurate predictions, we must go to a higher order in the cross coupling coefficients. This is likely to also shift the predicted values of the neutrino mixing angles and would be  interesting to study, which is left for future work.

\subsection{Neutrino Parameters}\label{nuparam}

\begin{figure}
    \centering
    \includegraphics[width=\linewidth]{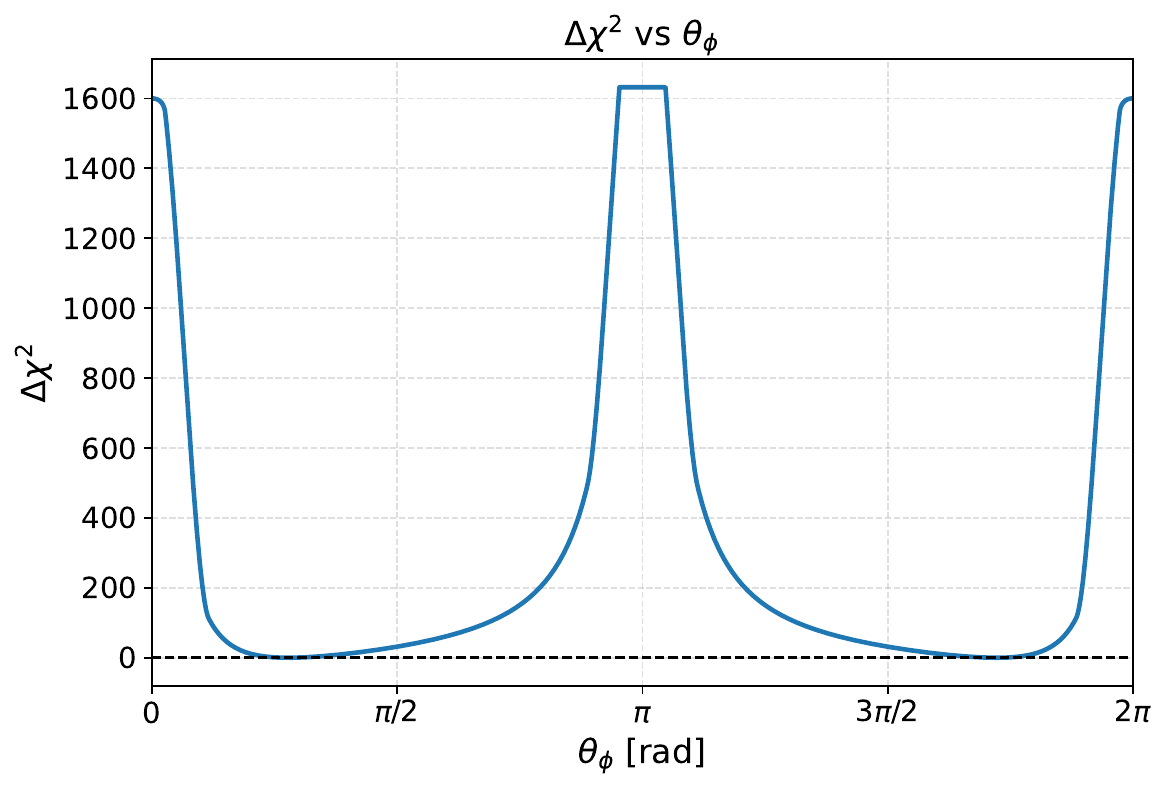}
    \caption{The $\chi^2$ difference vs $\theta_{\phi}$, showing the regions of best fit phases. We performed this $\chi^2$ fit to $\sin^2\,\theta_{13}$ and $\sin^2\,\theta_{23}$ to predict $\sin^2\,\theta_{12}$ from our derived sum rule Eq. \ref{sumrule}. Note the two values, the second valley corresponding to $\theta_{\phi}\approx298.5^{\circ}$, gives a more favorable range of $\sin^2\,\theta_{12} $.}
    \label{fig:preferredregion}
\end{figure}

Fitting the values reported in the $3\sigma$ range of values of $\sin^2\,\theta_{13}$ and $\sin^2\,\theta_{23}$ by NuFIT~\cite{Esteban:2024eli}, we predict $\theta_{\phi}\approx298.5^{\circ}$ as shown in Fig. \ref{fig:preferredregion}, $\sin^2\,\theta_{13}\approx0.024$, $\sin^2\,\theta_{23}\approx0.581$, $\sin^2\,\theta_{12}\approx0.297$, and a Dirac CP violating phase of $\delta_{CP}\approx282.5^{\circ}$. It is interesting that our prediction for the CP violating phase is so close to maximal, and our prediction of $\sin^2\,\theta_{12}$ differs from the recent results of JUNO~\cite{JUNO:2025gmd}. However, it still produces mixing angles consistent with data at the $3\sigma$ level~\cite{Esteban:2024eli,deSalas:2020pgw}, which can perhaps become more precise with the addition of more terms from assigning $+1$ charge to both $\phi$ and $\chi$ under the $\mathbb{Z}_4$ symmetry, or by incorporating the RGE running of the mixing angles in the case of nearly degenerate neutrino masses, both of which are left to future work. We fit our flavor parameters to the mass squared differences reported by NuFIT~\cite{Esteban:2024eli} and got $\Delta m_{21}^2\approx7.5\times10^{-5}\;\text{eV}^2$ and $\Delta m_{31}^2\approx2.45\times10^{-3}\;\text{eV}^2$, favoring normal ordering, and a symmetry breaking scale of around $v_{\chi}, v_{\phi} \sim \mathcal{O}(10 \text{TeV})$ corresponding to $y_D \sim \mathcal{O}(10^{-6})$ implying a low-scale seesaw mechanism. In principle, one can achieve a high-scale seesaw mechanism by looking at a symmetry breaking scale of $v_{\chi}, v_{\phi} \sim \mathcal{O}(10^{11} \text{TeV})$, however this would produce gravitational waves ruled out by the Planck bound~\cite{Pagano:2015hma} with too high frequency, as outlined in Sec. \ref{highseesaw}. Further, we find our flavon potential parameters to be $(\mu_{\phi}^2, \; \mu_\chi^2) \sim \mathcal{O}(100 \; \text{TeV}^2)$, $(f_i, \,  \tilde{f}_i) \sim \mathcal{O}(1)$, $\varepsilon_{1,2,3}\sim \mathcal{O}(0.1)$, and $\varepsilon_4v\sim \mathcal{O}(0.1 \; \text{TeV})$, where $v$ is an energy scale comparable to the flavor symmetry breaking scale.

\section{Gravitational Wave Spectrum from Domain Walls}\label{Section 6}
It is known that biased domain walls annihilate to produce gravitational waves. One can get the spectrum of the gravitational waves emitted at time $t$ as a function of frequency $f$, by solving 
\begin{equation}
    \Omega h^2(f,t)=\frac{h^2}{\rho_c(t)}\frac{d\rho_{GW}(t)}{d \ln f} \; .
\end{equation}
Red-shifting to present day, we can evaluate the peak amplitude and the peak frequency~\cite{Gelmini:2020bqg}
\begin{align}
    \Omega h^2|_{peak} \approx10^{-67} \frac{f_{\sigma}^4}{\varepsilon_b^2}\left( \frac{v}{\text{TeV}}\right )^4 \; ,\\
    f_{peak} \approx 3\times10^3 \text{Hz}\left(\frac{\varepsilon_b v}{f_{\sigma}\text{TeV}}\right)^{\frac{1}{2}} \; ,
\end{align}
where $\varepsilon_b$ was introduced in Eq. \ref{eq:6}. To plot the gravitational wave spectrum, we assumed a broken power law that obeys $\Omega h^2\propto f^3$ for $f<f_{peak}$ and $\Omega h^2 \propto f^{-1}$ for $f> f_{peak}$, motivated by numerical studies of domain walls~\cite{Hiramatsu:2013qaa}. The flavor dependence of the gravitational wave spectrum lies in $\varepsilon_b$ introduced in Eq. \ref{eq:6}, which in turn $V_b$ is related to $\varepsilon_{\phi}$ and $\varepsilon_{\chi}$ defined in Eq. \ref{eq:19} and \ref{eq:20} depending on whether we are analyzing $\mathbb{Z}_2$ or $\mathbb{Z}_3$ domain walls. Specifically, $\varepsilon_\phi$, is a function of the quartic couplings: $\varepsilon_2,\,f_2,\,f_3,\,f_1$, and the cubic coupling $f_4$; while $\varepsilon_\chi$ is a function of the quartic couplings: $\varepsilon_3,\,\tilde{f}_3,\,\tilde{f_2},\,\tilde{f_3}$, and the cubic coupling $\varepsilon_4$.

\subsection{Gravitational Waves from High-Scale See-Saw}\label{highseesaw}
As discussed in Sec. \ref{nuparam}, having a symmetry breaking scale of $\mathcal{O}(10^{11}\text{TeV})$, is achievable, but would produce either gravitational waves with amplitude ruled out by PLANCK~\cite{Pagano:2015hma} or with too high a frequency. A symmetry breaking scale of $10^{11}\text{TeV}$ corresponds to an $\varepsilon_b \sim 10^{-3}$, which in turn corresponds to the bound on $f_\sigma$ of $10<f_{\sigma}<10^{33}$ from Eq. \ref{eq:10}. Taking the lower limit of $f_\sigma$, we calculate the peak amplitude 
\begin{equation}
    \Omega h^2|_{peak} \approx10^{-67} \frac{f_{\sigma}^4}{\varepsilon_b^2}\left( \frac{v}{\text{TeV}}\right )^4 \sim10^{-67}\times10^{-2}\times10^{44}\approx 10^{-25} \; ,
\end{equation} which is not ruled out by PLANCK, but would correspond to a peak frequency $f\approx 10^{12} \; \text{Hz}$ far outside sensitivities. 
In contrast, for smaller frequencies, we would require $f_\sigma \sim 10^{30}$, corresponding to a peak amplitude of $\Omega h^2|_{peak} \sim 10^{91}$, far outside the reported PLANCK bound. Hence for good gravitational wave predictions, our model prefers low scale seesaw which we go into next. 
We would also like to point out that our model favors normal ordering for neutrinos purely from the $\chi^2$ minimization. We cannot gather this information from the production of gravitational waves as the mass squared differences are too similar between normal and inverted hierarchy to produce noticeable gravitational wave deviations. Additionally, the most meaningful parameter affecting the gravitational wave spectrum in the flavor sector is $\varepsilon_\phi$, due to its dependence on the VEVS of our scalars. We can see how this relates to the mixing angles see, eq.~\ref{angles}, which would primarily be affected through the VEVs of the scalar fields. Hence we see that the VEVs cannot be too different between normal and inverted orderings to produce similar angles since the deviations between normal and inverted ordering in the reactor angle $\theta_{13}$ are so small(see~\cite{Gariazzo:2018pei} for precise values), and hence the spectrum of gravitational waves won't change much.
\subsection{Gravitational Waves from Low-Scale See-Saw}
\begin{figure}
    \centering
    \includegraphics[width=\linewidth]{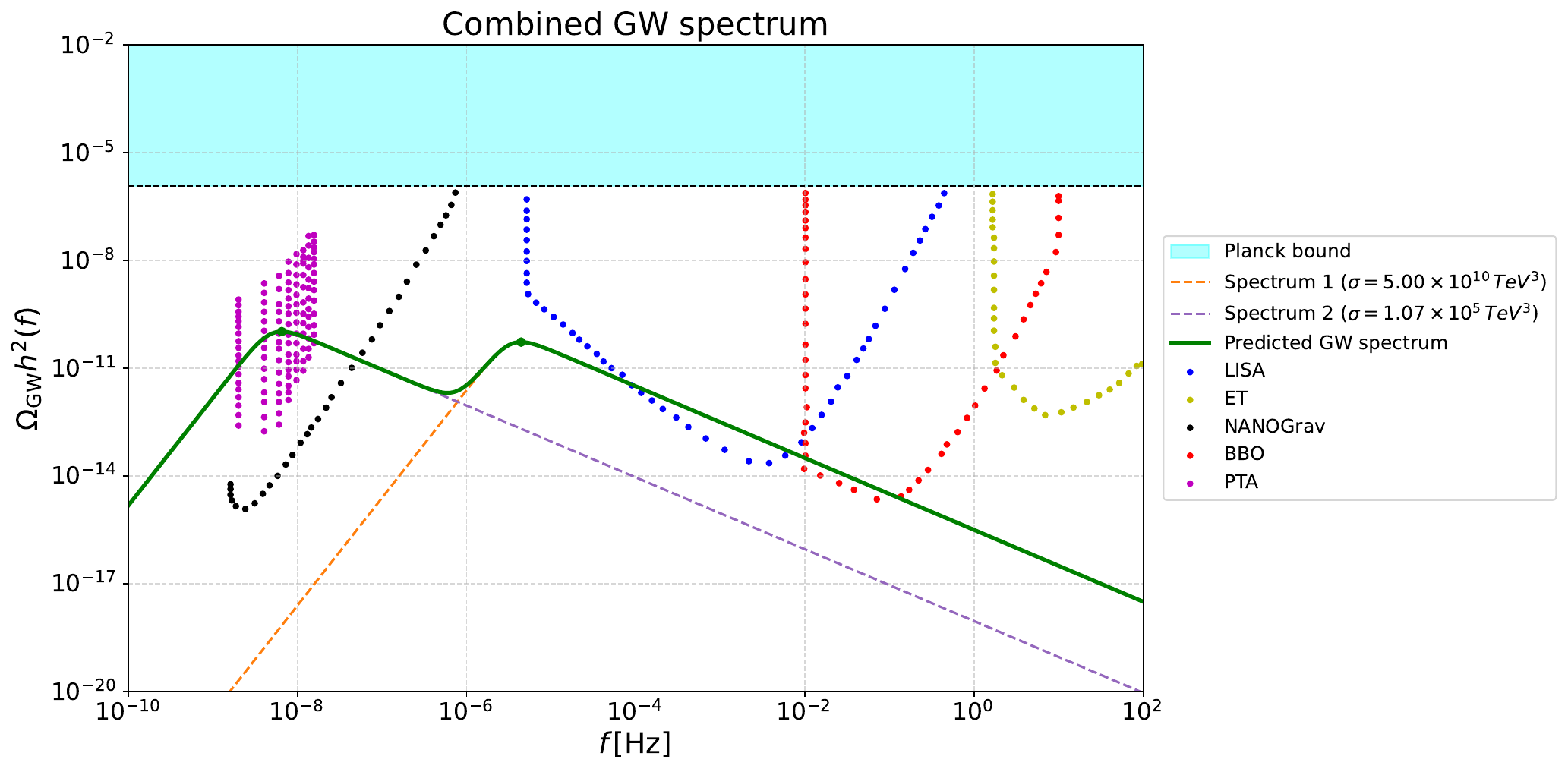}
    \caption{Gravitational-wave spectrum from domain wall annihilation with calculated $v_{\phi}\approx30~\text{TeV}, v_{\chi} \approx10$ TeV with appropriate domain wall tension. We see this could produce a GW signal that interestingly intersects with the observed Pulsar Timing Array signal.}
    \label{fig:gw_spectrum}
\end{figure}

We assume a smooth broken power law~\cite{Notari:2025kqq}, and plot in Fig.~\ref{fig:gw_spectrum} the combined Gravitational wave spectrum as a function of frequency using parameters found in the calculation of the mixing angles. We see that the combined spectrum results in signals that can be potentially detectable in near future gravitational wave experiments. 

\begin{figure}[t]
    \centering
    \includegraphics[width=\linewidth]{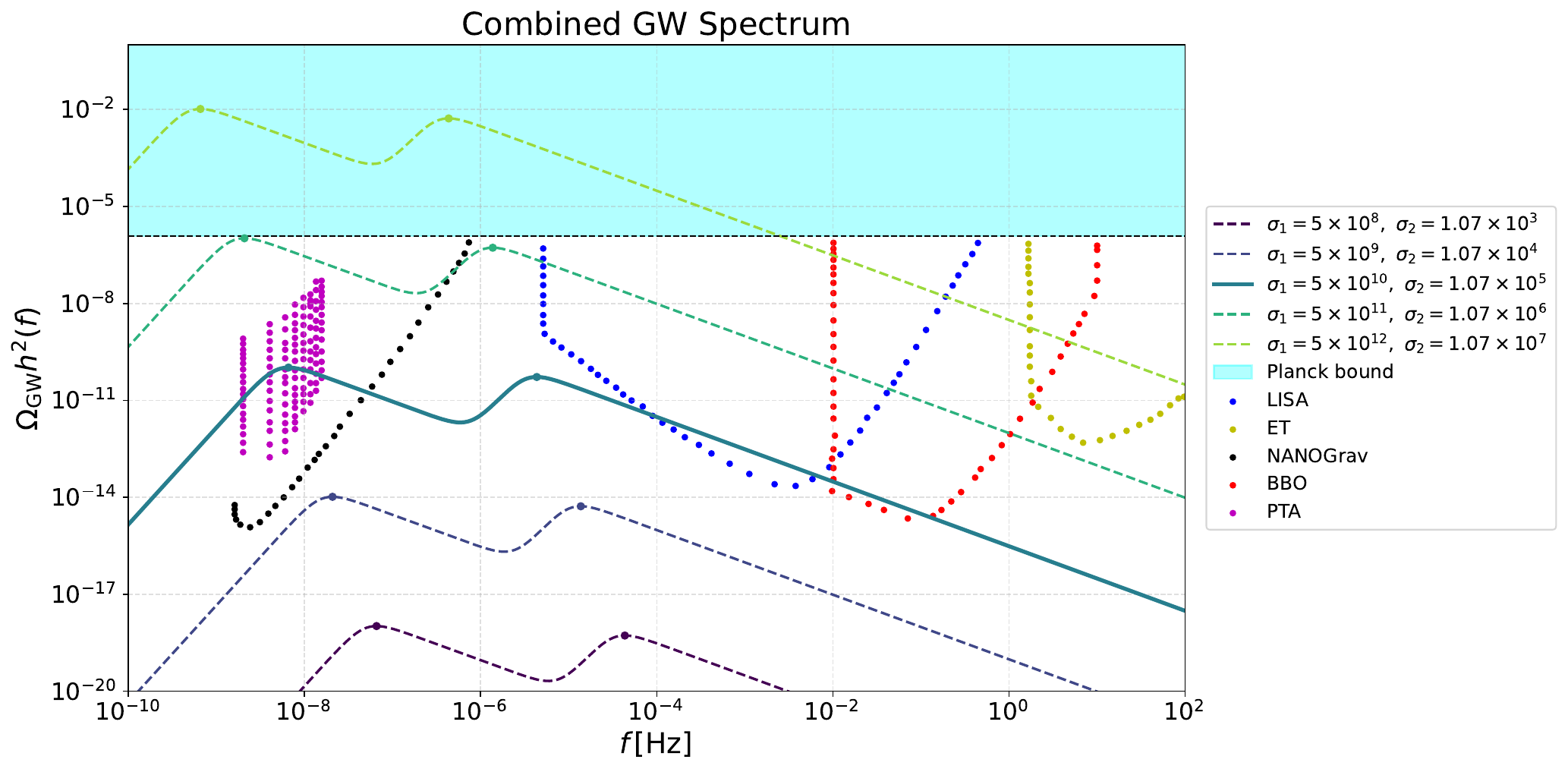}
    \caption{Gravitational Wave amplitude dependence on domain wall tension. We see that the greater the domain wall tension, the greater the amplitude of the gravitational waves which is intuitively expected. Note, the signal intersecting the Pulsar Timing Array signal~\cite{NANOGrav:2023gor} corresponds to a tension that gives a ratio in the bounds of Eq. \ref{eq:10}.}
    \label{fig:tension_dependence}
\end{figure}

The dependence of the combined spectra on the domain wall tension $\sigma$ is interesting because it dictates the detectability of gravitational waves caused by the annihilation of the domain walls. This is intuitively expected, as a larger tension would indicate a higher energy density of the domain walls, hence resulting in an increase in amplitude of the spectra, as can be seen in Fig. \ref{fig:tension_dependence}. We note that a domain wall tension of $\sigma=5\times10^{10}~\text{TeV}^3$ with a bias parameter of $\varepsilon_b=10^{-13}$ for the $\mathbb{Z}_3$ domain walls corresponding to the flavon field $\phi$ and $\sigma=1.07\times10^5 \text{TeV}^3$ with a bias parameter of $\varepsilon_b=2.14\times10^{-22}$ for the $\mathbb{Z}_2$ domain walls corresponding to the flavon field $\chi$ gives a signal intersecting with the observed Pulsar Timing Array signal, corresponding to a flavor symmetry breaking scale of $\mathcal{O}(10)$ TeV. Such a tension corresponds to a $\frac{\varepsilon_b}{f_\sigma}$ ratio (calculated from best fit values in Sec. \ref{Section 5}) within our previously derived bounds in Eq. \ref{eq:10}, hence the observed signal could be the result of the annihilation of cosmologically viable domain walls. However, the expressions dictating the annihilation of domain walls have been derived in the context of those that form from the spontaneous breaking of a $\mathbb{Z}_2$ symmetry, so we assume it follows similarly in our case. The precise relations for the $A_4$ symmetry case would require detailed numerical simulations on the evolution and dynamics of domain walls. However, we do not suspect serious deviations to the spectral shape of the gravitational waves produced by the domain walls other than perhaps slight shifts in the spectral slopes~\cite{Notari:2025kqq}. So we expect our results are accurate but further analysis is left to future work.

\par
We see that there is a direct link between the bias of the domain walls and flavor physics. This provides insight into the cosmological implications of high energy BSM physics, giving us information complementing particle collider experiments. Recent works have discussed flavon cross couplings giving rise to biases~\cite{Fu:2025qhf}, but we have given analytical expressions demonstrating the relations between the cross couplings and the biases.  

\subsection{Signal-To-Noise Ratio}
\begin{figure}
    \centering
    \includegraphics[width=\linewidth]{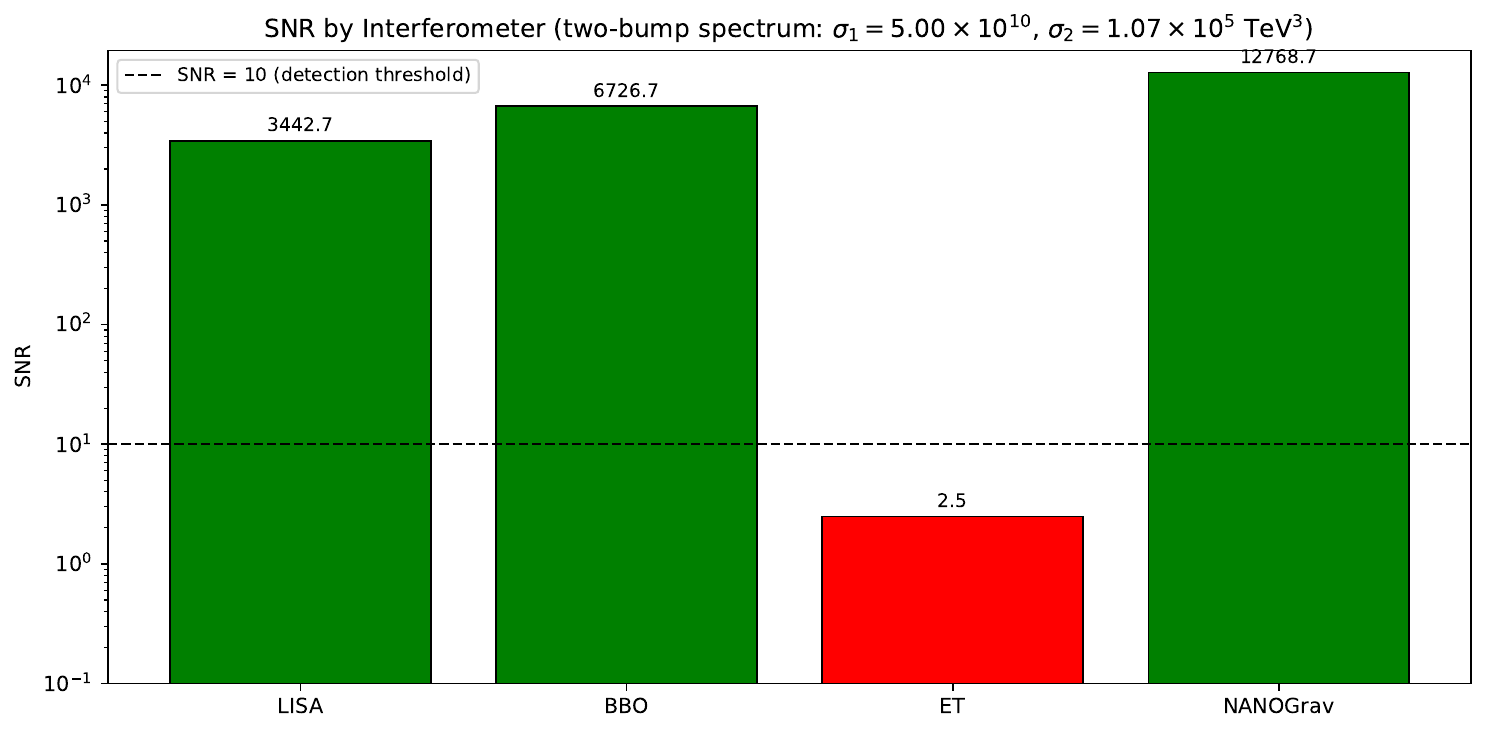}
    \caption{The SNR for a domain wall tension of $\sigma =5\times10^{10}\text{TeV}^3$. We note that $\text{SNR}>10$ for LISA, BBO, and NANOGrav.}
    \label{fig:snr_calc}
\end{figure}

The signal-to-noise ratio(SNR) is an important aspect to consider when discussing the sensitivity curves of gravitational wave experiments. Although there may be overlap of a gravitational wave spectrum with certain experiments such as Fig. \ref{fig:gw_spectrum}, doesn't mean it will be a meaningful detection. We have to introduce the idea of SNR to quantify whether or not a signal is meaningful and can correspond to a detection, with a $\text{SNR} \geq10$ qualifying a detection. We won't go into the specific details regarding the calculation of the SNR, but there is ample literature on the subject e.g.~\cite{Allen:1997ad,Kudoh:2005as,Thrane:2013oya,Caprini:2019pxz,Brzeminski:2022haa}. The SNR is defined to be 
\begin{equation}
    \text{SNR}\equiv 
    \sqrt{2\text{t}_{\text{obs}}\int_{f_{\text{min}}}^{f_{\text{max}}} df\left( \frac{h^2\Omega_{\text{GW}}(f)}{h^2\Omega_{\text{EXP}}(f)}\right)^2}
\end{equation}
where $\Omega_{\text{GW}}$ is our calculated values and $\Omega_{\text{EXP}}$ is the experimental noise sensitivity. We take $\text{t}_{\text{obs}}$ to be 4 years and $f_{\text{min}}$ and $f_{\text{max}}$ are the minimum and maximum frequency ranges for the experiments respectively. We calculate this for the our benchmark tensions of $\sigma = 5\times10^{10}\text{TeV}^3$ and $\sigma=1.07\times 10^5\text{TeV}^3$note that this tension gives us a detection for LISA, BBO, and NANOGrav, as well as matching the observed PTA signal, but is not detectable by ET as seen in Fig. \ref{fig:snr_calc} and \ref{fig:gw_spectrum}.

\subsection{Double Peak Feature}

Further, we would like to comment on the double peak in our gravitational wave spectrum. This double peak is an artifact of two distinct domain wall classes, $\mathbb{Z}_2$ and $\mathbb{Z}_3$, annihilating. This is a feature that has been noted for domain walls produced by non-abelian discrete symmetries ~\cite{Gelmini:2020bqg, Fu:2024jhu,Fu:2025qhf,Jueid:2023cgp}. However, double peaks show up in the context of current-carrying domain walls ~\cite{Ghoshal:2025gci}, as well as first order phase transitions ~\cite{Jana:2025vyb,Paul:2020wbz}; so a brief discussion is warranted. Our case of domain walls is distinguishable form the previous two scenarios by looking at the frequencies the peaks are and the relative maximum values of the two peaks. In the current-carrying domain wall case, we see that the two peaks occur at very different amplitudes whereas in our case, they are comparable. In the case of first order phase transitions, the peaks occur at much higher frequencies than the case with domain walls. Thus the two situations are distinguishable from our case by looking at what value of frequencies the peaks are, how separated they are, and the relative maximum values of the two peaks.

\subsection{Isocurvature Perturbations: GW Contributions and PBH Formation}

Domain walls can produce isocurvature perturbations which are also potential sources of gravitational waves. We proceed to determine the contribution to gravitational waves due to isocurvature perturbations. We estimate the isocurvature induced gravitational waves using the radiation era, one-scale domain wall framework in~\cite{Lu:2024sigw}. The induced gravitational wave contribution due to isocurvature is given by
 \begin{equation}
\Omega_{{\rm ind},i}(k)
=
\Omega_{{\rm GW},\;i}^{\mathrm{(before)}}(k)
+\Omega_{{\rm GW},\;i}^{\mathrm{(after,\;res)}}(k)
+\Omega_{{\rm GW},\;i}^{\mathrm{(after,\;LV)}}(k),
\qquad i=\phi,\chi\;,
\label{eq:induced_gw_components}
\end{equation}
where $\Omega_{{\rm GW},\;i}^{\mathrm{(before)}}(k)$ are the gravitational waves induced by scalar perturbations before DW annihilation, and $\Omega_{{\rm GW},\;i}^{\mathrm{(after,\;res)}}(k)$ is the resonantly enhanced contribution to GW after annihilation. The term $\Omega_{{\rm GW},\;i}^{\mathrm{(after,\;LV)}}(k)$ corresponds to the GW induced from large $\nicefrac{k}{q}$ modes where $k$ is the momentum of the induced GW and $q$ is the momentum of the scalar perturbations sourcing the GW after annihilation.  The wave numbers $k_{a,i}$ and the normalizations $B_i$ of these quantities are given in terms of model parameters 
\begin{equation}
k_{a,i}\simeq
4\pi A
\left(
\frac{\epsilon_{b,i}v_i^4}
{2\eta_0\sigma_i}
\right)^{1/2},
\qquad
B_i\simeq
\frac{2^{3/2}\pi^{5/2}}{3}
\frac{G^2\sigma_i^2}{\eta_0^2}\;.
\label{eq:induced_gw_model_inputs}
\end{equation}
Assuming formation temperatures $T_f \simeq v_i$, a network scaling coefficient of $A=1$, and $\eta_0 = 1.47 \times 10^8 \; \rm s $, the annihilation wavenumbers become $k_{a,\phi}\simeq4.14\times10^{-4}\,\mathrm{Hz}$ and $k_{a,\chi}\simeq6.26\times10^{-7}\,\mathrm{Hz}$. The wall energy fraction is given by 
 \begin{equation}
f_w(t_{\rm ann})
\simeq \frac{32\pi A G}{3}
\frac{\sigma^2}{\Delta V}
= \frac{32\pi A G}{3}
\frac{\sigma^2}{\epsilon_b v^4}\;.
\label{eq:dw_energy_fraction}
\end{equation}
For the $\mathbb{Z}_2$ and $\mathbb{Z}_3$ domain wall benchmarks, the wall energy fractions are $f_w\simeq5.36\times10^{-3}$ and $5.01\times10^{-3}$, respectively. The ratio of isocurvature induced GW to direct annihilation GW is given by 
\begin{equation}
R(f)=
\frac{
\Omega_{{\rm ind},\phi}(2\pi f)
+\Omega_{{\rm ind},\chi}(2\pi f)
}{
\Omega_{{\rm dir},\phi}(f)
+\Omega_{{\rm dir},\chi}(f)
}\;.
\label{eq:induced_direct_ratio}
\end{equation}
Maximizing the ratio across the frequency, we get a maximum of $8.3 \times 10^{-5}$ happening at $f \simeq 5.3 \times 10^{-5} \; \mathrm{Hz}$. Thus, the contribution to GW due to isocurvature perturbations induce negligible changes to the direct annihilation GW signature.

 In principle, CMB anisotropies can provide further constraints on isocurvature/scalar perturbations sourced by domain walls. However, the observed CMB anisotropies that constrain isocurvature range from $10^{-4}-10^{-1}\, \mathrm{Mpc}^{-1}$ ~\cite{Planck:2018jri} which correspond to frequencies of $10^{-19}-10^{-16} \, \mathrm{Hz}$ . The peak of the gravitational waves generated by the domain walls in our scenario centers around $10^{-10}-10^2 \, \mathrm{Hz}$ . The domain wall annihilation event is thus associated with comoving scales of $k_{\mathrm{ann}} \sim 10^5 - 10^{17} \mathrm{Mpc}^{-1}$. Even if domain walls source scalar perturbations, they do so at scales much smaller than the ones constrained by CMB data.

 Further, if the density of domain walls is large near the time of annihilation, primordial blackholes(PBH) can be formed~\cite{Passaglia:2021jla}. To make sure that PBH formation is significantly suppressed in our setup, we look at the variance in energy density defined as $\sigma_w = \frac{f_w}{\sqrt{N}}$, not to be confused with the domain wall surface tension $\sigma$, in~\cite{Lu:2024ngi} with a value of $\sigma_w \sim \mathcal{O}(10^{-1})$ indicating enhancements of PBH formation. However in our case we are safely below the threshold of $\sigma_w \sim \mathcal{O}(10^{-1})$ where
 \begin{equation}
     \sigma_w = \frac{f_w}{\sqrt{N}}=\frac{32\pi G \sigma t_{\text{ann}}}{\sqrt{N}}\approx \frac{10^{-33}\text{TeV}^{-2}\times10^{20}\text{TeV}^6}{10^{-12}\text{TeV}^4\times \sqrt{N}}\sim\mathcal{O}(10^{-3})
 \end{equation} where $N=4$ for $\mathbb{Z}_3 $ and $N=6$ for $\mathbb{Z}_2$ and we take $t_{\text{ann}}=\frac{\sigma}{\Delta V}$, so the production of PBH from isocurvature is subdominant, with $\sigma _w \approx 2\times10^{-3}$.

\section{Conclusion}\label{sec:conclusion}

In this paper, we proposed a new model invariant under $A_4 \times \mathbb{Z}_4$, and discussed both the particle and cosmological implications of the model. We consider two flavons $\phi$ and $\chi$, under which only $\chi$ carries a negative $\mathbb{Z}_4$ charge. This is important because it allows the existence of cubic couplings in the flavon potential. This cubic coupling in a singular flavon potential produces a bias in the $\mathbb{Z}_3$ invariant solutions, splitting the solution classes into two subclasses with different energies. Hence, it partially lifts the degeneracy of vacuum states. Furthermore, due to the $\mathbb{Z}_4$ charge assignments chosen, it allowed for cubic couplings in the potential that couples both $\phi$ and $\chi$. This similarly produces a bias that lifts the degeneracy between the vacuum states. In our approach, the full flavon potential respects the $A_{4} \times \mathbb{Z}_4$ symmetry, as opposed to the previous studies which explicitly broke this symmetry to generate a bias between vacua.

Furthermore, we discussed our particle physics model in the context of our modified vevs including cubic couplings in both the individual potentials and the cross coupling potential. This also had interesting implications for the prediction of mixing angles, and CP violation. We found the best fit values for the parameters in our flavon potential by considering $\sin\,\theta_{12}$. We predict a CP violating phase of $\delta_{CP}\approx282.5^{\circ}$ and found that deviation from maximal CP violation is an effect that comes at second order in our expansion parameters $\varepsilon_\phi$ and $\varepsilon_\chi$. This further begs to question what other interesting phenomena has its origins in higher order effects, and is something that should be further looked into. We have also used the detectability of gravitational waves emitted by the collapse of domain walls to constrain the flavor symmetry breaking scale to $\mathcal{O}(10\text{TeV})$, leading to a model preferring a low scale seesaw mechanism. One can predict the gravitational wave spectrum produced for a high scale seesaw mechanism $\mathcal{O}(10^{11}\text{TeV})$, but the gravitational waves produced would have either too high a frequency or too big an amplitude to be detected as discussed in Sec. \ref{highseesaw}. 

Given our best fit parameters, we are able to calculate the energy bias between sets of $\mathbb{Z}_2$ domain walls and sets of $\mathbb{Z}_3$ domain walls. We find, with appropriate choice of wall tensions, cosmologically safe domain walls that will collapse and preserve nucleosynthesis~\cite{Gelmini:2020bqg}. We then calculate the spectra of gravitational waves due to the collapse of domain walls. We find that for a tension of $\sigma\sim10^{10}~\text{TeV}^3$, the gravitational wave signatures can be detected in near future gravitational wave observatories. Interestingly, the spectrum also corresponds to a signal that was observed by the Pulsar Timing Array~\cite{NANOGrav:2023gor}, hence providing an interesting possible explanation for such a signal. 

 Our model thus predicts naturally unstable domain walls, which provides an explanation for the annihilation of domain walls without the addition of ad hoc explicit symmetry breaking terms. The cubic and cross terms in the flavon potential give rise to a deviation from TBM mixing, rendering realistic predictions for neutrino mixing parameters and observable gravitational wave signals. Our model thus demonstrates the close relations between particle physics and cosmology that should be further investigated.

We summarize the main results of our paper: 
\begin{itemize}
    \item We included all renormalizable terms allowed by the symmetry in the scalar potential inducing a bias without the need to explicitly break the symmetry.
    \item We have a unique model that produces realistic neutrino mixing angles, CP violating phase, and neutrino mass squared differences favoring Normal Ordering.
    \item We find that observable gravitational waves from the annihilation of domain walls requires the flavor symmetry breaking scale to be $v_{\phi,\chi}\sim \mathcal{O}(10\text{TeV})$, hence favoring a low scale seesaw mechanism.
    \item Given our parameter scan, we found a region of parameter space yielding gravitational waves potentially observable in near future gravitational wave detectors, coinciding with the observed Pulsar Timing Array Signal. 
\end{itemize}

\section*{Acknowledgments}
We thank Anish Ghoshal for bringing to our attention this interesting topic and for his very useful discussions. CMS acknowledges the support of the U.S. National Science Foundation (NSF) Graduate Research Fellowship Program (GRFP) and the Eugene Cota-Robles Fellowship provided by the University of California, Irvine. HM acknowledges the support of the U.S. Department of Education GAANN Fellowship. The work of M.-C.C. was partially supported by U.S. NSF under grant number PHY-2210283. 

%%%%%%%%%%%
%%%%%%%%%%%

\appendix 
\refstepcounter{section}
%\section{Appendix A} \label{appendix}
\section*{Appendix A: $A_{4}$ Product Rules}
\label{appendixA}
Here, we list the various tensor products of the different representations of $A_4$, There are four representations $\textbf{r}$: \textbf{1}, \textbf{1}', \textbf{1}'', and \textbf{3}. With the following tensor product decompositions, 
\begin{align*}
     \textbf{1} \times\textbf{r}=\textbf{r},\quad  \textbf{1} \times \textbf{1}''=\textbf{1}'',\quad \textbf{1}' \times\textbf{1}'=\textbf{1}'',\quad \textbf{1}''\times\textbf{1}''=\textbf{1}',\\ \textbf{1}' \times\textbf{3}=\textbf{1}'' \times \textbf{3}=\textbf{3}\quad, \textbf{3}\times\textbf{3}=\textbf{1}+\textbf{1}'+\textbf{1}''+\textbf{3}_S +\textbf{3}_A \; .
\end{align*}

In the AF basis, the tensor product of two triplets of $A_4$ contract as follows:
\begin{alignat}{2}
   & (ab)_{\mathbf{1}} &&= a_1 b_1 +  a_2 b_3 + a_3 b_2 , \\
   & (ab)_{\mathbf{1}'} &&= a_3 b_3 + a_1 b_2 + a_2 b_1 , \\
   & (ab)_{\mathbf{1}''} &&= a_2 b_2 + a_3 b_1 + a_1 b_3 , \\
   & (ab)_{\mathbf{3}_S} &&= \frac{1}{2} ( 2 a_1 b_1 - a_2 b_3 - a_3 b_2, 2 a_3 b_3 - a_1 b_2 - a_2 b_1, 2 a_2 b_2 - a_3 b_1 - a_1 b_3 )^T \\
   & (ab)_{\mathbf{3}_A} &&= \frac{1}{2} (  a_2 b_3 - a_3 b_2, a_1 b_2 - a_2 b_1, a_3 b_1 - a_1 b_3 )^T \;.
\end{alignat}
In the MR basis, we have 
\begin{alignat}{2}
   & (ab)_{\mathbf{1}} &&= a_1 b_1 + a_2 b_2 + a_3 b_3 , \\
   & (ab)_{\mathbf{1}'} &&= a_1 b_1 +  \omega a_2 b_2 + \omega^2 a_3 b_3 , \\
   & (ab)_{\mathbf{1}''} &&= a_1 b_1 +  \omega^2 a_2 b_2 + \omega a_3 b_3 ,  \\
   &  (ab)_{\mathbf{3}_S} &&= \frac{\sqrt{3}
    }{2} (  a_2 b_3 + a_3 b_2,  a_3 b_1 + a_1 b_3 ,   a_1 b_2 + a_2 b_1)^T \\
   & (ab)_{\mathbf{3}_A} &&= \frac{i}{2} (  a_2 b_3 - a_3 b_2,a_3 b_1 - a_1 b_3 ,  a_1 b_2 - a_2 b_1)^T \;.
\end{alignat}

%\section{Appendix B} \label{appendix2}
\refstepcounter{section}
\section*{Appendix B: Solutions to Flavon Cubic Potential}
\label{appendix2}
We list the solutions to the cubic potential in the Altarelli-Feruglio basis, which is the basis in which we calculate the corrections due to cross couplings for the potentials shown in Eq. \ref{equation18} and Eq. \ref{equation19}.
\begin{align}
   \langle\chi\rangle= \left \{ \begin{pmatrix}
    1\\1\\1
    \end{pmatrix},\begin{pmatrix}
    1\\\omega\\\omega^2
    \end{pmatrix},\begin{pmatrix}
    1\\\omega^2\\\omega
    \end{pmatrix}
    \right \}\frac{v_{\chi}}{\sqrt{3}},
\end{align} and 
\begin{align}
   \langle \phi \rangle_-= \left \{ \begin{pmatrix}
    \sqrt{3}\\0\\0
    \end{pmatrix},\begin{pmatrix}
    -\frac{1}{\sqrt{3}}\\\frac{2}{\sqrt{3}}\omega^2\\\frac{2}{\sqrt{3}}\omega
    \end{pmatrix},\begin{pmatrix}
     -\frac{1}{\sqrt{3}}\\\frac{2}{\sqrt{3}}\omega\\\frac{2}{\sqrt{3}}\omega^2
    \end{pmatrix},\begin{pmatrix}
     -\frac{1}{\sqrt{3}}\\\frac{2}{\sqrt{3}}\\\frac{2}{\sqrt{3}}
    \end{pmatrix}
    \right \}v_{\phi-},
\end{align}
and
\begin{align}
\langle \phi \rangle_+=\left \{ \begin{pmatrix}
    -\sqrt{3}\\0\\0
    \end{pmatrix},\begin{pmatrix}
    \frac{1}{\sqrt{3}}\\-\frac{2}{\sqrt{3}}\omega^2\\-\frac{2}{\sqrt{3}}\omega
    \end{pmatrix},\begin{pmatrix}
     \frac{1}{\sqrt{3}}\\-\frac{2}{\sqrt{3}}\omega\\-\frac{2}{\sqrt{3}}\omega^2
    \end{pmatrix},\begin{pmatrix}
     \frac{1}{\sqrt{3}}\\-\frac{2}{\sqrt{3}}\\-\frac{2}{\sqrt{3}}
    \end{pmatrix} \right \}v_{\phi+}
\end{align}
for the $\mathbb{Z}_2$ and the two $\mathbb{Z}_3$ preserving vacua respectively, where $\omega=e^{\frac{2\pi i}{3}}$. 

% Bibliography

%% [A] Recommended: using JHEP.bst file
%% \bibliographystyle{JHEP}
%% \bibliography{biblio.bib}

%% or
%% [B] Manual formatting (see below)
%% (i) We suggest to always provide author, title and journal data or doi:
%% in short all the informations that clearly identify a document.
%% (ii) please avoid comments such as "For a review'', "For some examples",
%% "and references therein" or move them in the text. In general, please leave only references in the bibliography and move all
%% accessory text in footnotes.
%% (iii) Also, please have only one work for each \bibitem.

\bibliographystyle{JHEP}     
\bibliography{References}     

\end{document}